\documentclass[journal]{IEEEtran}

\usepackage[english]{babel}
\usepackage[utf8]{inputenc}
\usepackage{amsmath,amssymb,bm,mathtools}
\usepackage{booktabs,multirow}
\usepackage{tabularx}
\usepackage{xcolor}
\usepackage{graphicx}
\usepackage{cite}
\usepackage{microtype}
\usepackage{xparse}
\let\newblock\relax
\NewDocumentCommand{\newblock}{g}{\IfNoValueTF{#1}{\relax}{#1}}
\NewDocumentCommand{\newedit}{+g}{\IfNoValueTF{#1}{\relax}{#1}}
\NewDocumentCommand{\vtwo}{+g}{\IfNoValueTF{#1}{\relax}{#1}}
\graphicspath{{figures/}{./}}
\begin{document}

\title{\newblock{Geometric} Time-Domain Identification of Three-Phase Load Equivalents from Terminal Measurements}

\author{Francisco G. Montoya, Francisco de Leon, \IEEEmembership{Fellow, IEEE},
Francisco M. Arrabal-Campos, and Alfredo Alcayde%
\thanks{F. G. Montoya, F. M. Arrabal-Campos, and A. Alcayde are with the Department of Engineering, University of Almeria, Spain. F. de Leon is with the Department of Electrical and Computer Engineering, New York University, USA.}}

\maketitle

\begin{abstract}
This paper presents a \newblock{geometric} time-domain method for identifying three-phase load
equivalents from instantaneous voltage and current measurements at the point of
common coupling. \newblock{Measured waveforms are interpreted as trajectories in
Euclidean signal spaces, and load-equivalent parameters are recovered from the
geometry of those trajectories.} The method extends a previously published single-phase
\newblock{geometric identification formulation} to three- and four-wire systems
and places special emphasis on the three-wire case, where no neutral voltage is measured and the
terminal data must satisfy coupled Kirchhoff constraints. The main advance over
the earlier analytical formulation is a sampled-data implementation based on
local time windows, normalized matrix equations, harmonic-projection derivative
and primitive coordinates, explicit geometric identifiability tests, passivity
constraints, and energy/Kirchhoff residuals.
The method does not force a model when the measured trajectory lacks enough
information; instead, it reports low-rank or ill-conditioned windows as
low-confidence evidence. Numerical
\newedit{simulations} with clean data, measurement noise, window-length sweeps, and sensor
delay show that the method accurately identifies informative three-phase
trajectories and exposes structurally degenerate cases such as pure
single-frequency excitation for higher-order three-wire models.
\vtwo{For a given admissible topology the identified circuit closes the
instantaneous terminal energy balance of the measured load over the analysis
window.}
\end{abstract}


\section{Introduction}

\IEEEPARstart{M}{odern} power systems are increasingly measured, controlled,
and protected through digital platforms. In this context, equivalent
\newblock{load and network-section parameters} are no longer only planning data; they
can become real-time state information for protection, monitoring, converter
control, and adaptive operation \cite{arif2017load,milanovic2012international}.
Most measurement-based load-identification methods estimate aggregate static or
dynamic parameters from events, disturbance records, phasor-domain quantities,
or optimization models
\cite{renmu2006composite,knyazkin2004parameter,wang2017robust}. Those
approaches are essential for system-level studies, but they are not designed to
recover an element-level energetic equivalent directly from the instantaneous
terminal waveforms of \newblock{a measured load or feeder section}.

A different school of thinking starts from the time-domain geometry of the measured
waveforms. Geometric algebra provides a compact language for subspaces,
oriented hypervolumes, and differential-geometric constructions
\cite{hestenes2012clifford,macdonald2010linear}. \newblock{In power-system
waveform applications}, this view is closely related to identifying
\newblock{physical load-equivalent elements} from
trajectory geometry rather than from a phasor fit or a spectral power
decomposition \cite{hong2015lissajous}. The single-phase \newblock{geometric
identification method published in 2021} formalized this idea by constructing a
waveform space vector we coined spacor, whose
osculating subspace yields closed-form expressions for \(R\), \(L\), and \(C\)
parameters \cite{montoyaTPWRD1}. \newblock{That earlier study} was mainly theoretical and
analytical: it proved that the parameters can be computed from terminal
time-domain data when the required geometric information is present.

Subsequent experimental validation showed that the single-phase method can
be operated with laboratory measurements, provided that numerical
differentiation, acquisition noise, and finite records are treated methodologically
\cite{arrabal2025experimental}. The present paper takes the next step. It
extends the identification problem to three-phase equivalents, with special
attention to three-wire systems where no neutral potential is available. It
also introduces the sampled-data layer needed to make the theory operational in
this setting: finite windows, matrix equations, window-local derivative and
integral coordinates, normalization, and diagnostic screening.

The three-phase extension is not a simple repetition of three independent
single-phase identifications. In four-wire systems, phase-to-neutral voltages
make a phase-wise formulation possible. In three-wire systems, however, branch
variables are hidden behind line-to-line voltages and constrained line
currents. Delta parallel and wye series equivalents may also become difficult
to distinguish if the local waveform trajectory has insufficient rank. A
practical method must therefore identify parameters and, at the same time,
quantify whether the measured window contains enough information to support the
chosen physical equivalent.

This paper concentrates on the identification problem. \newblock{It does not
attempt to introduce a complete power theory; instead, it uses energy
conservation as an internal physical validation criterion for the identified
equivalent.} The identification itself is not accepted only because a set of
parameters fits the terminal equations. A candidate
equivalent must also close the instantaneous energy balance implied by its
resistors, inductors, and capacitors. This energy-admissibility criterion is a
new practical layer of the method and is used here as a physical validation
gate for every processed window.

\newblock{The manuscript develops a unified formulation for
three-phase wye and delta equivalents in three- and four-wire systems, and then
turns the published geometric construction into a windowed matrix estimator
that can be applied directly to sampled terminal records.} The implementation
\newblock{adds a derivative and primitive reconstruction layer based on
harmonic projection, FIR filters, polynomial fits, or regularized finite
differences, avoiding high-order pointwise differentiation.} It also
\newblock{makes identifiability explicit through the geometric rank and
conditioning of the local time-domain trajectory. It adds the
energy-admissibility test so that a numerically fitted
equivalent is accepted only when it closes the terminal energy balance.}
\newblock{Because each window reduces to a fixed-topology least-squares
problem, the computational cost remains small and explicit. The numerical
evidence is then built from analytically controlled cases and realistic MATLAB
\newedit{simulations} with
unbalanced multisine feeders, measurement noise, gain and offset errors,
quantization, moving-average acquisition effects, window-length sweeps,
coherent current-sensor delay, and low-information cases that must be screened
out rather than over-fitted.}

\newedit{Taken together, these tests give a consistent quantitative picture. In
	the controlled three-wire delta and wye cases the windowed estimator reproduces
	the analytical branch constants and closes the terminal power balance at machine
	precision, with the residual rising only to the injected noise floor (of order
	\(10^{-3}\)) under 60 dB measurement noise. In the realistic 20 kHz multisine
	simulations, practical windows of 0.5--2 cycles keep the recovered branch \(R\),
	\(L\), and \(C\) values within a 5\% engineering band under clean data, 60 and
	40 dB noise, 12-bit quantization, and gain mismatch. The energy-admissibility
	gate accepts fifteen of the twenty-one practical-window configurations and
	rejects exactly the six affected by a 50 \(\mu\)s current-sensor delay, where
	the parameter error grows to about 37\%. The method therefore recovers physical
	parameters under ordinary measurement imperfections and, just as important,
	flags a coherent timing bias instead of disguising it as a reliable equivalent.}

\vtwo{The practical value of this method is that it determines an equivalent
circuit which, for the assumed admissible topology, closes the terminal energy
balance of the measured load over the analysis window. The energy residual is an
admissibility check for the identified terminal equivalent, not a claim that
terminal data uniquely determine the physical internal energy localization. The
identified \(R\), \(L\), and \(C\) elements therefore form a physically
structured equivalent rather than a black-box fit.}

\vtwo{The specific contributions of this paper, relative to the single-phase
geometric method of~\cite{montoyaTPWRD1} and its experimental validation
in~\cite{arrabal2025experimental}, are as follows.
\begin{itemize}
\item A unified three-phase, three- and four-wire formulation of the geometric
load equivalent, covering both wye and delta connections under coupled Kirchhoff
constraints with no measured neutral voltage.
\item A topology-dependent observability analysis that identifies which branch
forms are directly recoverable and proves the structural singularity of the
series-delta and parallel-wye three-wire cases.
\item A windowed matrix estimator that realizes the geometric (osculating-subspace)
construction numerically over a finite sampled window, with harmonic-projection
derivative and primitive coordinates.
\item An explicit identifiability layer based on the geometric rank and
conditioning of the local time-domain trajectory, screening low-information
windows rather than over-fitting them.
\item An energy-admissibility gate that accepts an identified equivalent only
when it closes the instantaneous terminal energy balance over the analysis
window.
\end{itemize}
This paper addresses identification. The element-level energy accounting that
the identified equivalents enable is pursued in separate work~\cite{paper2arxiv}.}

\section{Three-Phase Load Equivalents and Identification Equations}

\begin{figure}[!t]
	\centering
	\includegraphics[width=\columnwidth]{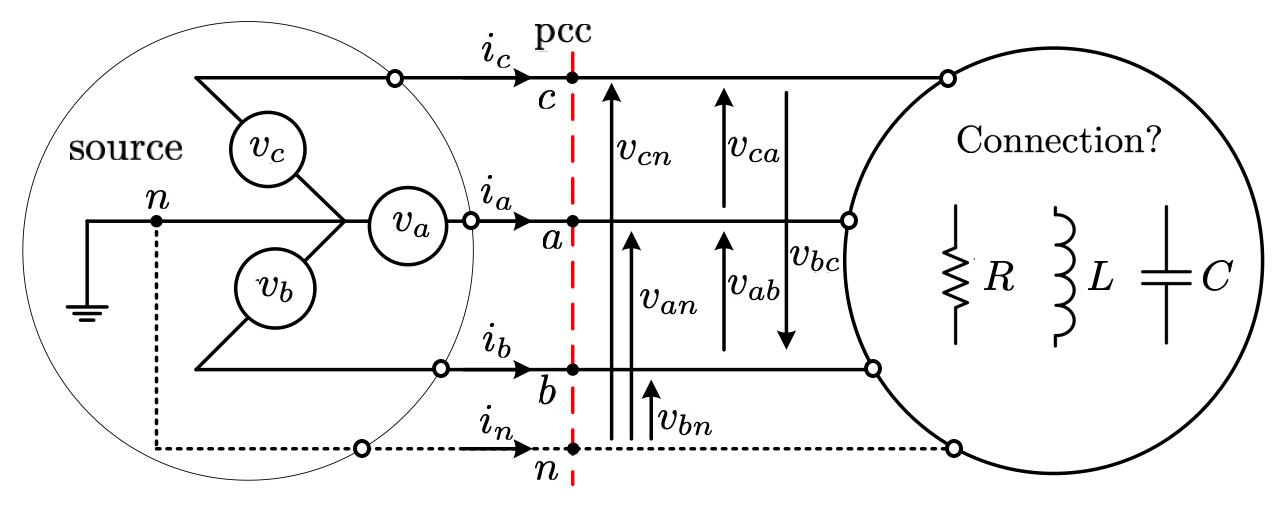}
	\caption{Three-phase load identification from terminal measurements at the PCC.}
	\label{fig:pcc_problem}
\end{figure}

The problem is to identify an equivalent three-phase load from measurements of
instantaneous voltages and currents at the point of common coupling (PCC), as
shown in Fig.~\ref{fig:pcc_problem}. The equivalent is required to reproduce
the measured terminal behavior and to provide a physically interpretable set of
parameters over the local observation window.

For a four-wire system, phase-to-neutral voltages are available or can be
directly measured. Each phase can therefore be treated as a single-phase
identification problem with a chosen branch topology. This problem was solved in \cite{montoyaTPWRD1} and experimentally validated in \cite{arrabal2025experimental}. For a three-wire system,
only line-to-line voltages are available and the line currents satisfy
\begin{equation}
	i_a+i_b+i_c=0,
	\qquad
	v_{ab}+v_{bc}+v_{ca}=0.
	\label{eq:threewire_constraints}
\end{equation}
The absence of a measured neutral makes the three-wire case a complex 
technical problem.

The identification process is as follows. First, a candidate topology is
selected: wye or delta, and series or parallel branches. Second, a space vector,
or \emph{spacor}, is built from the measured terminal variables and the
derivatives or integrals required by the chosen branch law. Third, KCL and KVL
are used to express the measured trajectory as a lower-dimensional object
embedded in a higher-dimensional Euclidean space. Finally, the multivector
associated with the model subspace is compared with the osculating multivector
obtained from the measured trajectory. In the symbolic version this comparison
produces closed-form quotients for
$R$, $L$, or $C$. In the sampled-data version used here,
the same equations are evaluated over finite windows and screened by rank,
conditioning, passivity, and residual diagnostics.

\subsection{Equivalent Model Families}

\begin{figure}[!t]
	\centering
	\includegraphics[width=0.72\columnwidth]{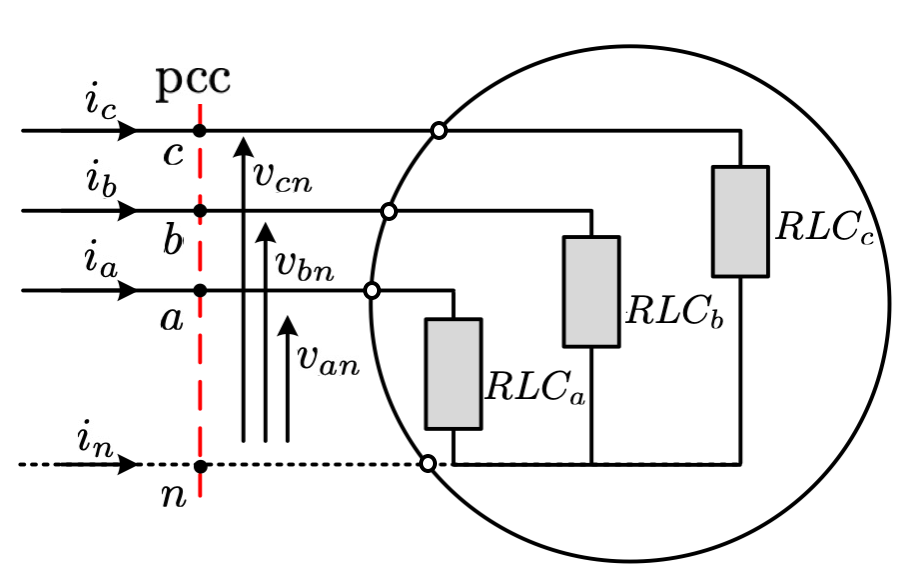}
	\caption{Wye-connected three-phase equivalent. In four-wire systems the neutral is measurable; in three-wire systems it may be virtual or absent.}
	\label{fig:wye_model}
\end{figure}

\begin{figure}[!t]
	\centering
	\includegraphics[width=0.82\columnwidth]{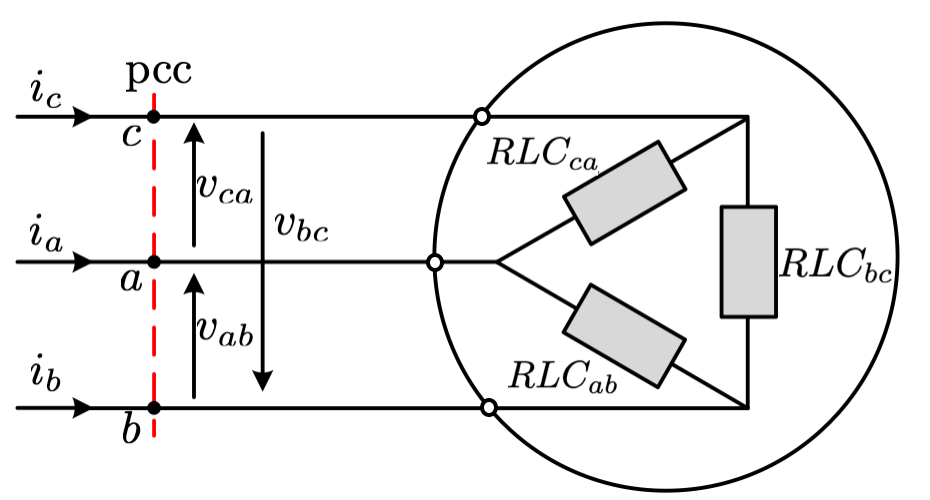}
	\caption{Delta-connected three-wire equivalent. Branch variables are inferred from line-to-line voltages and line currents.}
	\label{fig:delta_model}
\end{figure}

\begin{figure}[!t]
	\centering
	\includegraphics[width=\columnwidth]{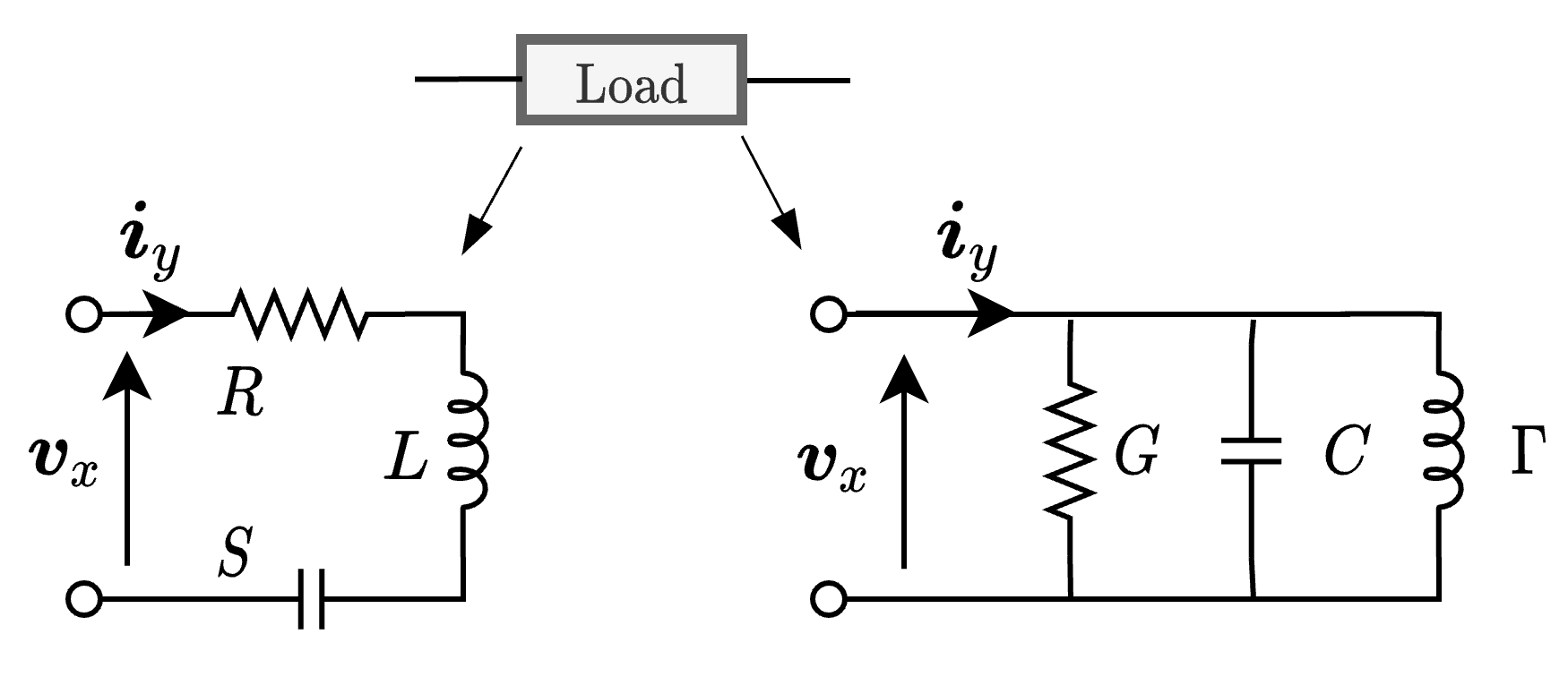}
	\caption{Series and parallel $RLC$ branch models used to build the three-phase library.}
	\label{fig:branch_models}
\end{figure}

The model library starts from the two basic three-phase connections shown in
Figs.~\ref{fig:wye_model} and \ref{fig:delta_model}. Each branch can be
represented by a series or parallel $RLC$ equivalent, as in
Fig.~\ref{fig:branch_models}. Simpler submodels are obtained by suppressing
terms that are not required or not identifiable in a local window.

\subsubsection{Four-Wire Wye Model}

When phase voltages $v_{an}$, $v_{bn}$, and $v_{cn}$ are available, a parallel
branch model can be written as
\begin{equation}
	i_x(t)=G_x v_{xn}(t)+\varGamma_x \breve v_{xn}(t)+C_x v_{xn}'(t),
	\qquad x\in\{a,b,c\},
	\label{eq:fourwire_parallel}
\end{equation}
where $v'$ denotes time derivative and $\breve v$  time integral of $v$, respectively, and
$\varGamma=1/L$. A series branch model is
\begin{equation}
	v_{xn}(t)=R_x i_x(t)+L_x i_x'(t)+S_x \breve\imath_x(t),
	\qquad S=1/C.
	\label{eq:fourwire_series}
\end{equation}
These equations are the phase-wise version of the single-phase method in
\cite{montoyaTPWRD1}. The four-wire case is included for completeness, while the
three-wire case is the focus of this paper.

\subsubsection{Three-Wire Delta Parallel Model}

\newblock{For a three-wire load, the directly resolvable delta family is the
parallel branch form. The reason is physical and geometric: the branch voltages
of a delta are exactly the measured line-to-line voltages, and the measured line
currents are obtained from KCL differences of the internal branch currents.}
For a delta parallel equivalent,
\begin{align}
	i_{ab} &= G_{ab}v_{ab}+\varGamma_{ab}\breve v_{ab}+C_{ab}v_{ab}',\\
	i_{bc} &= G_{bc}v_{bc}+\varGamma_{bc}\breve v_{bc}+C_{bc}v_{bc}',\\
	i_{ca} &= G_{ca}v_{ca}+\varGamma_{ca}\breve v_{ca}+C_{ca}v_{ca}'.
	\label{eq:delta_branch_model}
\end{align}
The measured currents satisfy
\begin{equation}
	i_a=i_{ab}-i_{ca},\qquad
	i_b=i_{bc}-i_{ab},\qquad
	i_c=i_{ca}-i_{bc}.
	\label{eq:delta_line_currents}
\end{equation}

\newblock{The full $G\varGamma C$ form is the model used by the sampled-data
implementation when the local trajectory has enough rank. To show the
geometric core without hiding it inside numerical linear algebra, the analytical
derivation is written here for the two-element $G\varGamma$ delta reference.}
The independent
measured variables can be chosen as
$v_{ab}$, $v_{bc}$, $i_a$, and $i_b$. The six-dimensional spacor is
\begin{equation}
	\bm y=
	v_{ab}\bm e_1+v_{bc}\bm e_2+\breve v_{ab}\bm e_3+
	\breve v_{bc}\bm e_4+i_a\bm e_5+i_b\bm e_6 .
	\label{eq:delta_waveform_vector_paper1}
\end{equation}
Using $v_{ca}=-(v_{ab}+v_{bc})$, KCL gives
\begin{align}
	i_a&=(G_{ab}+G_{ca})v_{ab}+G_{ca}v_{bc}
	+(\varGamma_{ab}+\varGamma_{ca})\breve v_{ab}
	+\varGamma_{ca}\breve v_{bc},\\
	i_b&=-G_{ab}v_{ab}+G_{bc}v_{bc}
	-\varGamma_{ab}\breve v_{ab}+\varGamma_{bc}\breve v_{bc}.
	\label{eq:delta_kcl_expanded_paper1}
\end{align}
Thus $\bm y$ lies in a four-dimensional subspace of the six-dimensional
measurement space. If
\begin{equation}
	\bm K=\bm a\wedge\bm b\wedge\bm c\wedge\bm d
\end{equation}
is the model hypervolume and $\bm K_{\mathrm{osc}}$ is the osculating
\newblock{hypervolume computed from trajectory derivatives through the wedge
product, then the identified parameters follow by comparing the scalar
coefficients of corresponding blades.} Specifically, if
\(\bm K_{\mathrm{osc}}=\sum_I s_I\bm e_I\), then \(s_{ijkl}\) denotes the
scalar coefficient multiplying
\(\bm e_i\wedge\bm e_j\wedge\bm e_k\wedge\bm e_l\). The quotients used for
the delta parallel \(G\varGamma\) model are
\begin{equation}
	\begin{aligned}
	G_{ab}&=\frac{s_{2346}}{s_{1234}},
	& G_{bc}&=\frac{s_{1346}}{s_{1234}},
	& G_{ca}&=\frac{s_{1345}}{s_{1234}},\\
	\varGamma_{ab}&=\frac{s_{1246}}{s_{1234}},
	& \varGamma_{bc}&=\frac{s_{1236}}{s_{1234}},
	& \varGamma_{ca}&=\frac{s_{1235}}{s_{1234}} .
	\end{aligned}
	\label{eq:delta_ga_quotients_paper1}
\end{equation}
\vtwo{The expanded symbolic forms of the \(s_I\) coefficients are provided in
the online derivation package at
\texttt{https://electrica.ual.es/spacor}. The quotients in
\eqref{eq:delta_ga_quotients_paper1} are exact: each branch parameter equals a
fixed ratio of blade coefficients of the osculating hypervolume, so the windowed
least squares used later does not discover an arbitrary equivalent---it
evaluates, over a finite record, the very linear identities that the geometric
construction makes exact. The matrix form is therefore a numerical
evaluation of \eqref{eq:delta_ga_quotients_paper1}, not a generic
regression onto KCL regressors.}

\subsubsection{Three-Wire Wye Series Model}

\newblock{The directly resolvable wye family in a three-wire load is the series
form. Here the measured line currents are the branch currents, with
\(i_c=-(i_a+i_b)\), while the unknown neutral voltage is eliminated by writing
KVL in terms of line-to-line voltages.} For a wye series equivalent without
measured neutral, the branch equations are
\begin{equation}
	v_{xn}=R_x i_x+L_xi_x'+S_x\breve\imath_x,\qquad x\in\{a,b,c\}.
	\label{eq:wye_series_branch}
\end{equation}
The measurable line voltages are
\begin{equation}
	v_{ab}=v_{an}-v_{bn},\quad
	v_{bc}=v_{bn}-v_{cn},\quad
	v_{ca}=v_{cn}-v_{an}.
	\label{eq:wye_line_voltage}
\end{equation}
For the two-element $RL$ series wye model, KVL gives the explicit measured
relations
\begin{align}
	v_{ab} &=
	R_a i_a+L_ai_a'-R_b i_b-L_bi_b',\\
	v_{bc} &=
	R_b i_b+L_bi_b'
	+R_c(i_a+i_b)+L_c(i_a'+i_b').
	\label{eq:wye_kvl_expanded_paper1}
\end{align}
The corresponding geometric construction uses the spacor
\begin{equation}
	\bm z=
	i_a\bm e_1+i_b\bm e_2+i_a'\bm e_3+
	i_b'\bm e_4+v_{ab}\bm e_5+v_{bc}\bm e_6 .
	\label{eq:wye_waveform_vector_paper1}
\end{equation}
Comparing the model hypervolume with the osculating hypervolume gives the same
type of \(s_I\)-coefficient quotients:
\begin{equation}
	\begin{aligned}
	R_a&=\frac{s_{2345}}{s_{1234}},
	& R_b&=\frac{s_{1345}}{s_{1234}},
	& R_c&=\frac{s_{2346}}{s_{1234}},\\
	L_a&=-\frac{s_{1245}}{s_{1234}},
	& L_b&=\frac{s_{1235}}{s_{1234}},
	& L_c&=\frac{s_{1246}}{s_{1234}} .
\end{aligned}
	\label{eq:wye_ga_quotients_paper1}
\end{equation}
\vtwo{As for the delta case, the expanded \(s_I\) expressions are provided in the
online derivation package and remain mechanically auditable.
The six branch parameters are recovered from the geometry of the measured
\((i_a,i_b,i_a',i_b',v_{ab},v_{bc})\) trajectory, not from a phasor
decomposition: \eqref{eq:wye_ga_quotients_paper1} is the same blade-coefficient
identity for the wye-series family, and the windowed estimator of
Section~\ref{sec:geom_matrix} simply evaluates it over a finite window.}

\begin{table*}[!t]
	\centering
	\caption{Topology-Dependent Identifiability from Terminal Data}
	\label{tab:identifiability}
	\footnotesize
	\setlength{\tabcolsep}{6pt}
	\renewcommand{\arraystretch}{1.15}
	\begin{tabularx}{\textwidth}{@{}l l l c >{\raggedright\arraybackslash}X@{}}
		\toprule
		\textbf{Connection / wires} & \textbf{Meas.\ vars} & \textbf{Branch form} & \textbf{Status} & \textbf{Reason} \\
		\midrule
		Wye, 4-wire & $v_{xn},i_x$ & $RLS$ series & I & Phase-to-neutral voltages measured; each phase reduces to single-phase~\cite{montoyaTPWRD1}. \\
		\addlinespace[1pt]
		Delta, 3-wire & $v_{xy},i_x$ & $G\varGamma C$ parallel & I & Branch voltages are the measured line voltages; branch currents from KCL of line currents. \\
		\addlinespace[1pt]
		Delta, 3-wire & $v_{xy},i_x$ & $RLS$ series & S & Needs internal branch currents; a circulating $h(t)$ leaves $i_x$ unchanged. \\
		\addlinespace[1pt]
		Wye, 3-wire & $v_{xy},i_x$ & $RLS$ series & I & Line currents are branch currents; neutral eliminated by KVL in $v_{xy}$. \\
		\addlinespace[1pt]
		Wye, 3-wire & $v_{xy},i_x$ & $G\varGamma C$ parallel & S & Needs absolute $v_{xn}$; a common-mode $u(t)$ leaves $v_{xy}$ unchanged. \\
		\addlinespace[1pt]
		Virtual neutral & $v_{xn'},i_x$ & $RLS$ series & C & Reference node $\sum v_{xn'}=0$ is a coordinate device, not a measured conductor; recovers branch sums. \\
		\bottomrule
	\end{tabularx}
	\\[2pt]
	\raggedright\footnotesize I: identifiable. S: structurally singular (observability singularity, not numerical failure). C: coordinate construction. $\varGamma=1/L$, $S=1/C$.
\end{table*}

\subsubsection{Structural Singularities and Virtual-Neutral Model}

\newblock{The two direct three-wire families above are not arbitrary choices.
They are the nonsingular combinations imposed by terminal observability. A
parallel wye equivalent would require the absolute phase-to-neutral voltages
\(v_{an}\), \(v_{bn}\), and \(v_{cn}\), but three-wire measurements determine
only their differences. Adding an arbitrary common-mode function \(u(t)\) to
all three phase potentials leaves \(v_{ab}\), \(v_{bc}\), and \(v_{ca}\)
unchanged while changing the branch voltages seen by a parallel wye model.
Thus the parameters are not unique without an additional neutral constraint.}
\newblock{Conversely, a series delta equivalent would require the internal
branch currents \(i_{ab}\), \(i_{bc}\), and \(i_{ca}\). Terminal currents give
only their differences; the transformation
\((i_{ab},i_{bc},i_{ca})\mapsto(i_{ab}+h,i_{bc}+h,i_{ca}+h)\) leaves
\(i_a\), \(i_b\), and \(i_c\) unchanged for any circulating component \(h(t)\).
Since a series branch law depends on the branch current itself, the associated
parameters are structurally indeterminate. These are observability
singularities of the three-wire measurement problem, not numerical failures.}

\newblock{A useful additional representation is obtained by introducing a
virtual neutral. This construction, related to the classical artificial-neutral
and \(n-1\) wattmeter viewpoints \cite{blondel1893measurement,depenbrock1980active,staudt2008fryze},
does not create a measured physical conductor. It selects a reference node that
allows a wye series equivalent to account for terminal behavior when the load is
better interpreted as a sum of line-to-line branches. The virtual-neutral
example in Section~\ref{sec:canonical_cases} shows that the identified wye
parameters recover the original line-to-line values through branch sums, which
is precisely the interpretation expected from the structural singularity of the
direct series-delta family.}

\vtwo{The singularities resolved here concern parameter observability:
within a chosen admissible family, the branch parameters either are or are not
recoverable from terminal data. They do not concern the uniqueness of the
internal energy localization across admissible families. Two families that
are both nonsingular here---for instance the delta parallel and the wye
series---can reproduce identical terminal data while localizing stored and
dissipated energy differently---a modeling question beyond the terminal
identifiability addressed here.}

\vtwo{Table~\ref{tab:identifiability} summarizes which branch form each connection
makes directly identifiable from its measured terminal variables, and which
combinations are structurally singular rather than merely ill-conditioned.}

\section{From Geometric Theory to Windowed Matrix Implementation}

\subsection{Geometric and Matrix Identification}
\label{sec:geom_matrix}

The \newblock{analytical geometric method published for the single-phase case}
constructs a time-dependent vector whose coordinates
are measured waveforms and their derivatives or integrals. Kirchhoff laws
confine this trajectory to a lower-dimensional geometric object.
\newblock{Equivalent parameters} are obtained by comparing the model subspace with the osculating
subspace of the measured trajectory.

\vtwo{In exact symbolic form, the geometric construction yields the closed
quotients \eqref{eq:delta_ga_quotients_paper1} and
\eqref{eq:wye_ga_quotients_paper1}, in which each branch parameter is a fixed
ratio of blade coefficients of the osculating hypervolume. In sampled data, those
same blade identities are evaluated over a short window on the order of a single
fundamental period---here 0.5--2 cycle windows, about 10--40~ms at 50~Hz---which
absorbs noise and lets the construction track slowly varying parameters. The
least-squares problem solved per window is not a generic regression; it is the
finite-window form of that identity.} For a window \(w\), the model is written as
\begin{equation}
	\mathbf y_w=\Phi_w\theta+\epsilon_w,
	\label{eq:generic_ls}
\end{equation}
where \(\mathbf y_w\in\mathbb R^{m_w}\) is the measured output vector,
\(\Phi_w\in\mathbb R^{m_w\times n_\theta}\) is built from the voltage/current
coordinates, derivatives, and integrals dictated by KCL or KVL, and
\(\theta\in\mathbb R^{n_\theta}\) collects the unknown branch parameters. The
integer \(n_\theta\) is therefore the number of estimator coordinates required
by the selected equivalent; for example, \(n_\theta=6\) for a three-wire delta
parallel \(G\varGamma\) model and \(n_\theta=9\) for a full delta parallel \(RLC\)
model written in \(G,\varGamma,C\) coordinates.

Before solving the least-squares problem, the matrix is scaled to remove the
arbitrary numerical effect of units. A full-parameter estimate is meaningful
only if the scaled local trajectory contains enough information:
\begin{equation}
	\operatorname{rank}_{\tau}(\tilde{\Phi}_w)=n_\theta,
	\qquad
	\kappa_2(\tilde{\Phi}_w)=
	\frac{\sigma_{\max}(\tilde{\Phi}_w)}
	{\sigma_{\min}(\tilde{\Phi}_w)}
	\le \kappa_{\max}.
	\label{eq:rank_condition}
\end{equation}
Here \(\operatorname{rank}_{\tau}\) is the numerical rank computed with a
tolerance \(\tau\), \(\sigma_{\max}\) and \(\sigma_{\min}\) are the extreme
singular values of the scaled regressor matrix, and \(\kappa_{\max}\) is a
user-defined acceptance threshold. In the MATLAB \newedit{simulations} below,
\(\kappa_{\max}=10^6\) is used as a conservative gate. \vtwo{Thus the geometric
quotients and the windowed least squares are two evaluations of the same linear
identities implied by Kirchhoff's laws and the chosen topology---symbolic on an
exact trajectory, numerical on a finite noisy record. The estimator does not
select an equivalent; it evaluates the equivalent that the geometry has already
fixed.}

\subsection{Identifiability and Information Content}

The method is not a frequency-domain identification method. Harmonics are not
the foundation of the theory. The relevant question is whether the time-domain
trajectory spans the geometric dimension required by the chosen equivalent.

For example, in a single-phase parallel $RLC$ model the regressors are
$v$, $\breve v$, and $v'$. A pure sinusoid satisfies
$v'=-\omega^2\breve v$, so the regressors cannot span a three-dimensional
space. The same principle explains the three-wire case: pure fundamental data
may be sufficient for a resistive model but insufficient for higher-order
unbalanced $RL$ or $RLC$ equivalents. Harmonics, switching, transients, ramps,
or modulation can all create the required time-domain rank.

This rank criterion is central to the implementation. Measurement noise is not
allowed to "repair" a structurally non-identifiable trajectory; such windows
are reported as low-information rather than used as high-confidence evidence.

\subsection{Energy-Admissibility Criterion}

Rank and conditioning answer an information question: Can the terminal trajectory support the number of unknowns of the selected equivalent circuit?. They
do not, by themselves, prove that the recovered parameters define a physically
admissible energy model. For this reason each accepted window is also tested by
a \vtwo{terminal energy-balance check (for the assumed topology)}.

Let \(p_{\mathrm{PCC}}(t)\) denote the instantaneous terminal power computed
from the measured voltage-current coordinates at the PCC. Once a candidate
delta or wye equivalent has been identified, the same window gives the
irreversible Joule term and the stored electromagnetic energy,
\begin{equation}
	p_{\mathrm d}(t)=
	\sum_{r\in\mathcal R} R_r i_r^2(t)
	=
	\sum_{g\in\mathcal G} G_g v_g^2(t),
	\label{eq:joule_term}
\end{equation}
\begin{equation}
	W(t)=
	\frac{1}{2}\sum_{\ell\in\mathcal L} L_\ell i_\ell^2(t)
	+
	\frac{1}{2}\sum_{c\in\mathcal C} C_c v_c^2(t).
	\label{eq:stored_energy_term}
\end{equation}
The energy residual is then
\begin{equation}
	\rho_E =
	\frac{
	\left\|
	p_{\mathrm{PCC}}(t)-p_{\mathrm d}(t)-W'(t)
	\right\|_2
	}{
	\left\|p_{\mathrm{PCC}}(t)\right\|_2+\varepsilon_E
	}.
	\label{eq:energy_admissibility_residual}
\end{equation}
Here \(\varepsilon_E\) only prevents division by zero in very low-power
windows. Equation~\eqref{eq:energy_admissibility_residual} is not a new power
definition. It is a conservation check: after the equivalent is identified, the
measured terminal power must be explainable as dissipated power plus the time
derivative of energy stored in the identified \(L\) and \(C\) elements
\vtwo{of the assumed equivalent}.

This criterion is important for two reasons. First, it prevents a numerically
well-conditioned least-squares fit from being interpreted as a physical
equivalent when the recovered elements do not close the energy balance.
Second, it separates random measurement noise from coherent measurement-chain
biases. Additive noise broadens \(\rho_E\) and the parameter distributions,
whereas a current-sensor delay produces a systematic phase error that may pass
some rank tests but fails the energy interpretation. Thus the method reports an
equivalent as high-confidence only when it is identifiable, passive,
Kirchhoff-consistent, and energy-admissible.

\subsection{Sampled-Data Numerical Implementation}

The published single-phase theory established the \newblock{geometric
construction} and its continuous-time \newblock{energy-equivalent interpretation}
\cite{montoyaTPWRD1}. The later experimental validation showed that the
single-phase idea can be operated with laboratory measurements
\cite{arrabal2025experimental}. The implementation used here adds a
different layer: it turns the exact geometric construction into a finite-window,
matrix-based, three-phase estimator with explicit signal-processing and
diagnostic steps. This layer is essential for three-wire systems because the
missing neutral, the KCL/KVL constraints, and the coexistence of several
energy-equivalent families make pointwise symbolic formulae unsuitable for
noisy sampled records.

The sampled-data implementation follows these steps:
\begin{enumerate}
	\item form overlapping local windows from the terminal voltage and current
	records;
	\item build the matrix pair $(\Phi,\mathbf y)$ required by the physical
	equivalent under test;
	\item compute derivative or primitive coordinates with a configured
	window-local signal model;
	\item scale the columns of $\Phi$ and the output rows so that conditioning is
	not dominated by units;
	\item solve the resulting least-squares problem and convert the estimator
	coordinates back to physical \(R\), \(L\), and \(C\) values;
	\item compute rank, condition number, coverage, passivity, KCL/KVL residuals,
	terminal-power residual, and the energy-admissibility residual
	\(\rho_E\).
\end{enumerate}

This is not merely a different coding style for the analytical formulae. The
windowed estimator changes how the method is used. Instead of differentiating a
single sample and trusting the resulting quotient, each estimate is supported by
all samples in the window and by independent consistency tests. In addition, the
same signal record can be tested against delta parallel RLC and wye series RLC
families without changing the physical measurements; only the KCL/KVL matrix
changes. If the local trajectory cannot support all parameters of a higher-order
equivalent, the corresponding rank and conditioning diagnostics make this
visible before the parameters are interpreted.

The derivative layer is also new with respect to the theoretical formulation.
The implementation avoids high-order pointwise numerical derivatives. The
three-wire \newedit{simulations} reported here require primitives or first derivatives
only. Derivative coordinates are obtained over the same observation window by a
configurable signal estimator: linear-phase FIR differentiators, local
polynomial filters, Tikhonov-smoothed finite differences, or harmonic
projection followed by analytical differentiation
\cite{paquelet2018fir,tseng2012differentiator}. The harmonic projection is the
default for the multisine simulations because it separates the physical
low-order spectral content used for excitation from broadband measurement
noise. The matrix/windowed procedure is summarized in
Appendix~\ref{app:windowed_implementation}.

For a fixed topology, the computational cost of one window is dominated by
forming $\Phi$ and solving a small least-squares problem. If the window has
$N$ samples and the model has $n_\theta$ unknowns, the cost is
$O(Nn_\theta^2+n_\theta^3)$, with $n_\theta\le 9$ for the three-wire families
used in this paper. Thus the method does not rely on a nonlinear search over a
large parameter space. Its practical limitation is not the algebraic cost but
the availability of informative windows and an appropriate physical equivalent
set for the measured feeder.

\section{Validation Strategy and Simulation \newedit{Studies}}

\subsection{Canonical Analytical Identification Cases}
\label{sec:canonical_cases}

The following analytical examples serve as controlled reference cases. Their
role is to demonstrate the \newblock{load-equivalent topologies} and the symbolic identification
path, while the MATLAB \newedit{simulations} in the next section test numerical robustness
under more realistic measurement conditions.

\subsubsection{Three-Wire Delta Parallel Case}

\begin{figure}[!t]
	\centering
	\includegraphics[width=0.82\columnwidth]{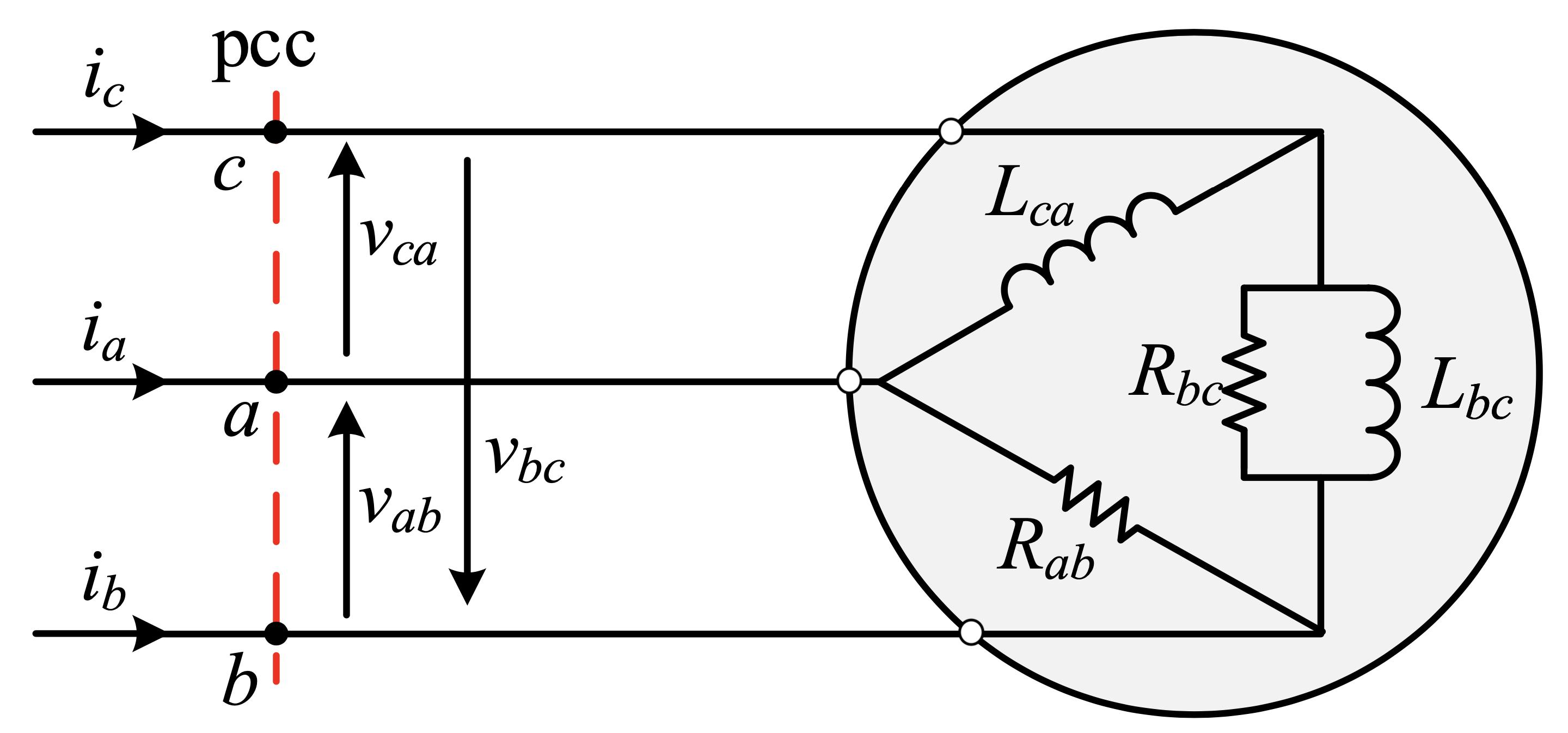}
	\caption{Three-wire unbalanced delta load with nonsinusoidal excitation.}
	\label{fig:delta_parallel_load}
\end{figure}

The delta example of Fig.~\ref{fig:delta_parallel_load} is the first core
three-wire case. The test load has unbalanced linear $RL$ branches:
$R_{ab}=2.0\,\Omega$, $R_{bc}=1.333\,\Omega$,
$L_{bc}=2\,\mathrm{mH}$, and $L_{ca}=3\,\mathrm{mH}$. Equivalently,
$G_{ab}=0.5\,\mathrm{S}$, $G_{bc}=0.750\,\mathrm{S}$,
$\varGamma_{bc}=500\,\mathrm{H}^{-1}$, and
$\varGamma_{ca}=333.3\,\mathrm{H}^{-1}$.

\begin{figure}[!t]
	\centering
	\includegraphics[width=\columnwidth]{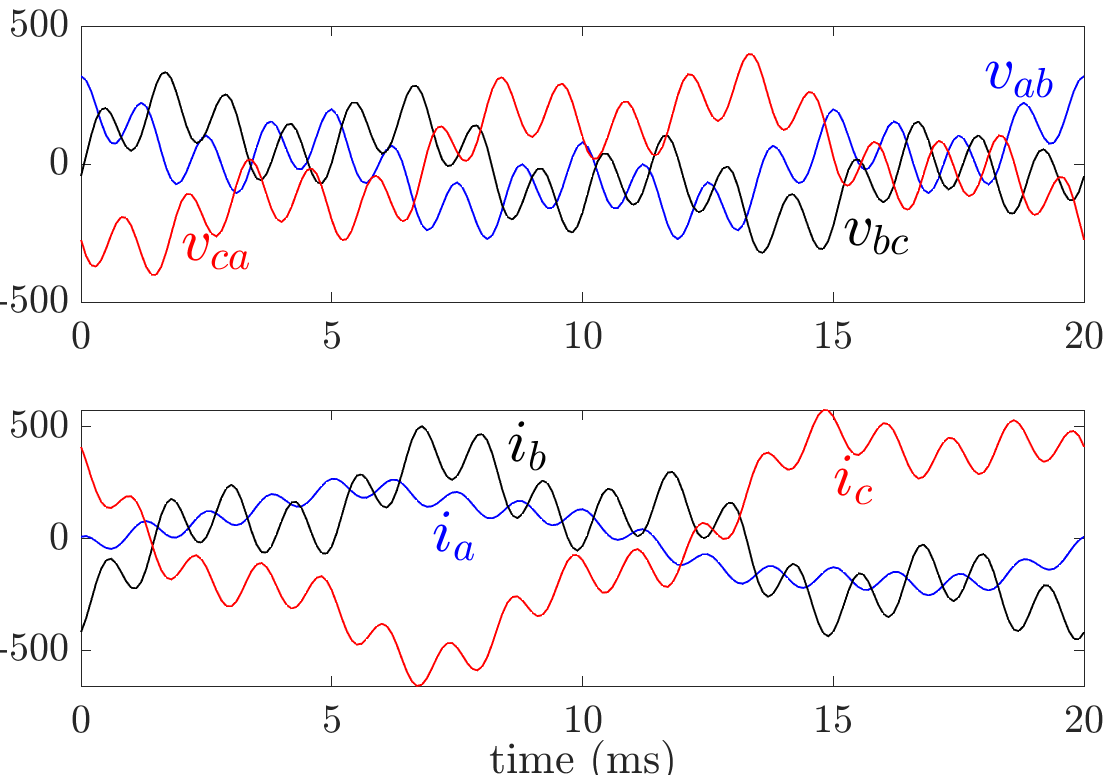}
	\caption{Voltage and current waveforms for the delta case.}
	\label{fig:delta_parallel_waveforms}
\end{figure}

\begin{figure}[!t]
	\centering
	\includegraphics[width=\columnwidth]{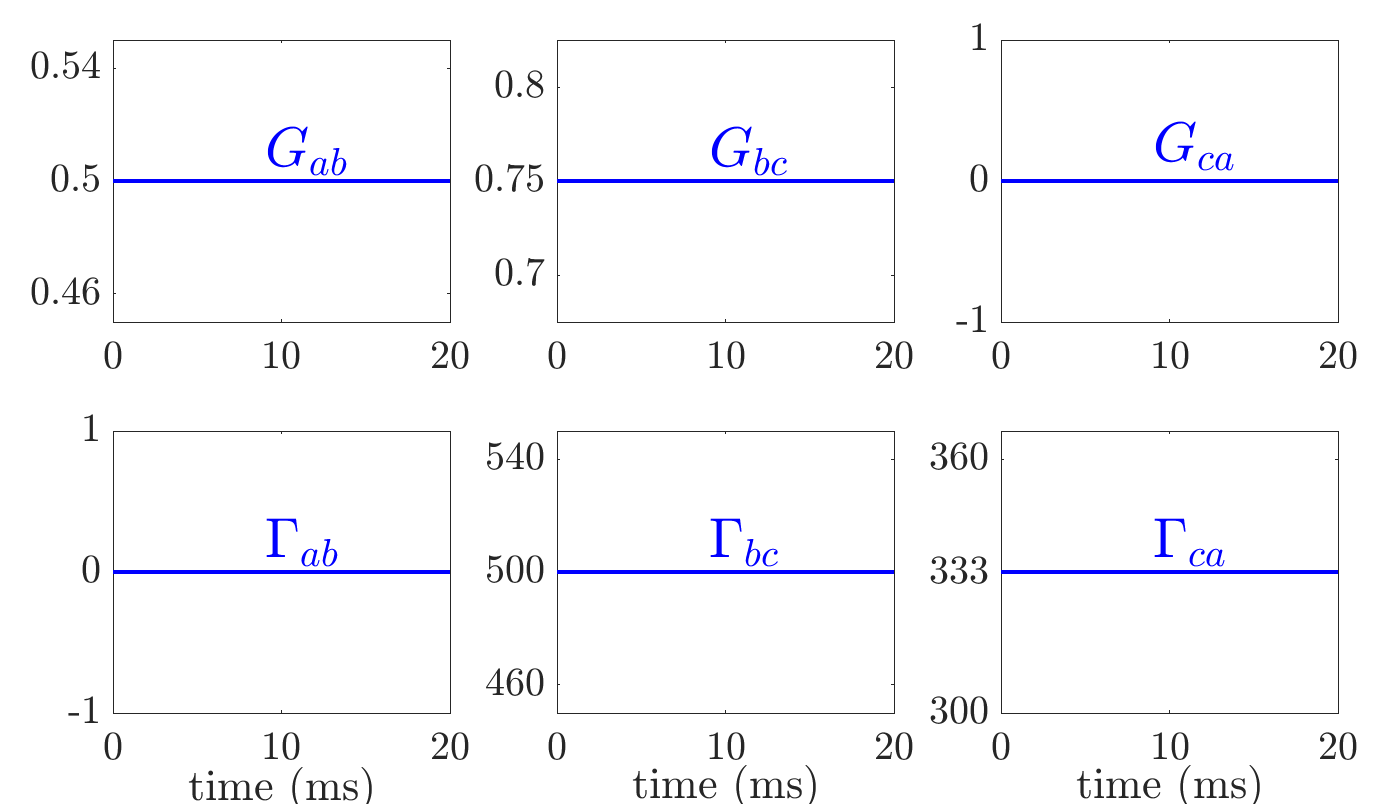}
	\caption{Delta identification result. This case is the clean reference for noise, window, and delay studies.}
	\label{fig:delta_parallel_identification}
\end{figure}

The line-to-line voltages used in this controlled test are
\begin{align}
	v_{ab}&=120\cos100\pi t+90\cos400\pi t+108\cos1600\pi t,\\
	v_{bc}&=132\cos(100\pi t-\pi/3)+100\cos(400\pi t-2\pi/3)\notag\\
	&\quad+120\cos(1600\pi t-2\pi/3),\\
	v_{ca}&=-(v_{ab}+v_{bc}).
\end{align}
These waveforms provide the geometric rank needed by the $G\varGamma$ delta
model, so the exact GA quotients and the windowed matrix estimator return the
same branch constants in the noise-free case. The waveforms and identified
parameters are shown in Figs.~\ref{fig:delta_parallel_waveforms}
and \ref{fig:delta_parallel_identification}.

\newedit{The terminal waveforms in Fig.~\ref{fig:delta_parallel_waveforms} are
strongly nonsinusoidal, so the local trajectory spans the full rank required by
the \(G\varGamma\) delta model. The identification is then exact: every window
in Fig.~\ref{fig:delta_parallel_identification} returns
\(G_{ab}=0.5\,\mathrm{S}\), \(G_{bc}=0.75\,\mathrm{S}\),
\(\varGamma_{bc}=500\,\mathrm{H}^{-1}\), and
\(\varGamma_{ca}=333.3\,\mathrm{H}^{-1}\) as flat lines with no scatter, and the
structural zeros \(\varGamma_{ab}=0\) (purely resistive \(ab\) branch) and
\(G_{ca}=0\) (purely inductive \(ca\) branch) are recovered without spurious
leakage. In this noise-free reference the windowed estimator matches the
analytical GA quotients to numerical precision.}

\subsubsection{Three-Wire Wye Series Case}

\begin{figure}[!t]
	\centering
	\includegraphics[width=0.8\columnwidth]{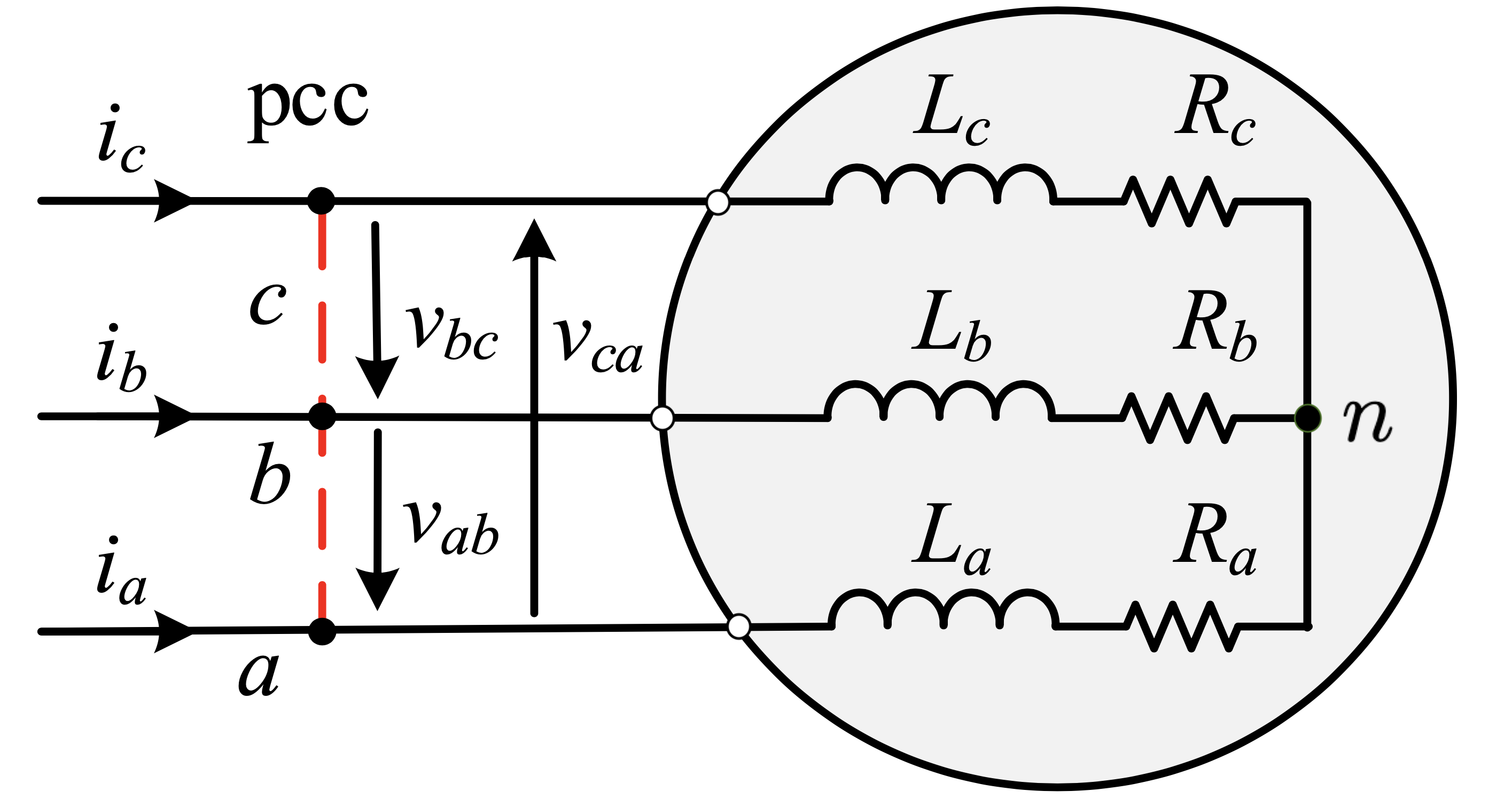}
	\caption{Unbalanced three-wire wye series load.}
	\label{fig:wye_series_load}
\end{figure}

The second core three-wire case is the series wye load of
Fig.~\ref{fig:wye_series_load}. The current-source excitation used in this
controlled case is
\begin{align}
	i_a&=120\cos\omega t+10\cos5\omega t,\\
	i_b&=240\cos(\omega t-2\pi/3)+10\cos(5\omega t-10\pi/3),\\
	i_c&=-i_a-i_b .
\end{align}
The branch parameters were
$R_a=0.5\,\Omega$, $R_b=0\,\Omega$, $R_c=7\,\Omega$,
$L_a=3\,\mathrm{mH}$, $L_b=7\,\mathrm{mH}$, and $L_c=0$.
This example is kept because it demonstrates that the three-wire wye problem
is solvable when formulated as a series branch problem driven by KVL.

\begin{figure}[!t]
	\centering
	\includegraphics[width=0.9\columnwidth]{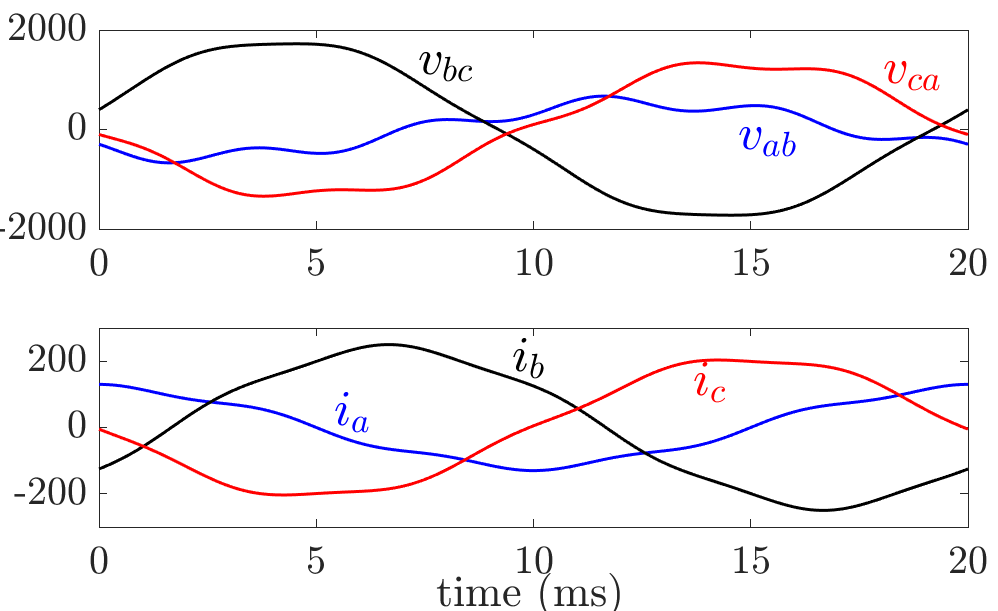}
	\caption{Line-to-line voltages and line currents for the wye series case.}
	\label{fig:wye_series_waveforms}
\end{figure}

\begin{figure}[!t]
	\centering
	\includegraphics[width=\columnwidth]{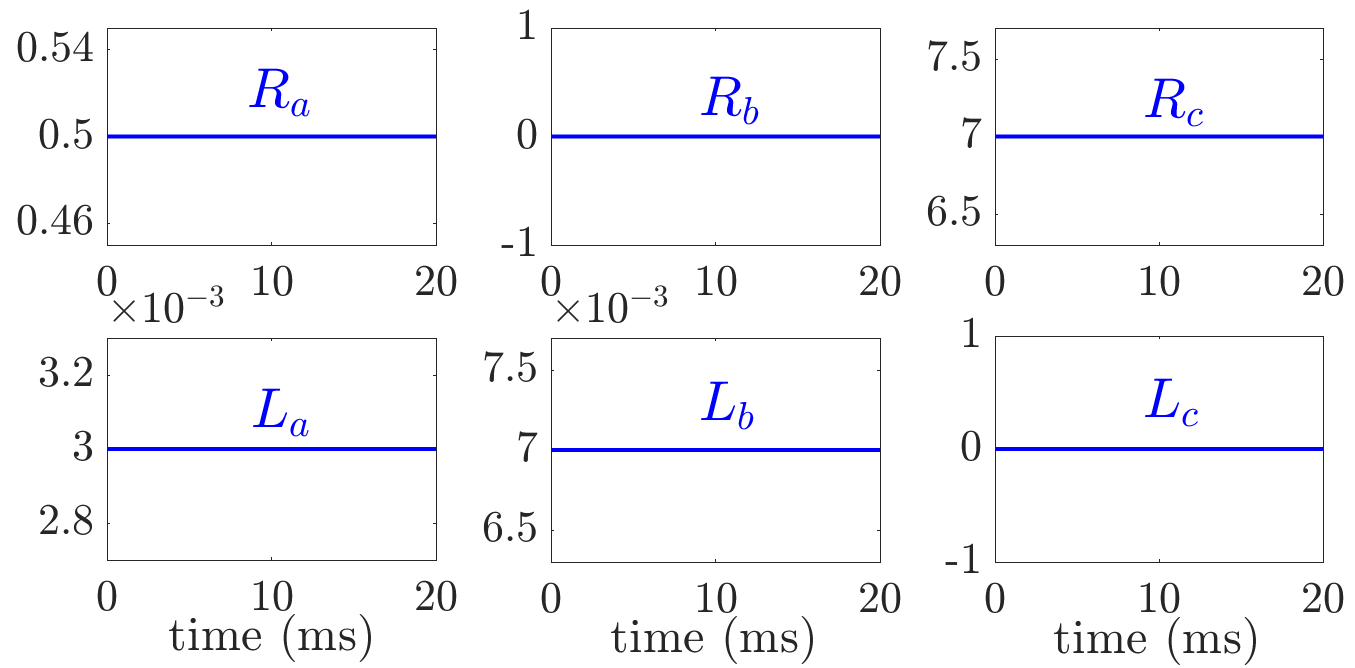}
	\caption{Wye series identification result, reported together with rank and residual diagnostics.}
	\label{fig:wye_series_identification}
\end{figure}

\subsubsection{Virtual Neutral Case}

The virtual-neutral construction is also preserved, but with a clearer role:
it is an identification coordinate system, not a hidden physical neutral. This
idea is connected with Blondel's polyphase measurement theorem
and with the Fryze-Buchholz-Depenbrock time-domain framework
\cite{blondel1893measurement,depenbrock1980active,staudt2008fryze}. From the
line voltages,
\begin{equation}
	v_{an'}=\frac{1}{3}(v_{ab}-v_{ca}),
	v_{bn'}=\frac{1}{3}(v_{bc}-v_{ab}),
	v_{cn'}=\frac{1}{3}(v_{ca}-v_{bc}),
	\label{eq:virtual_neutral_paper1}
\end{equation}
which ensures $v_{an'}+v_{bn'}+v_{cn'}=0$. This allows the use of
phase-wise single-phase identification with $i_{n'}=0$ when the chosen
equivalent is a virtual wye model.
\begin{figure}[!t]
	\centering
	\includegraphics[width=0.9\columnwidth]{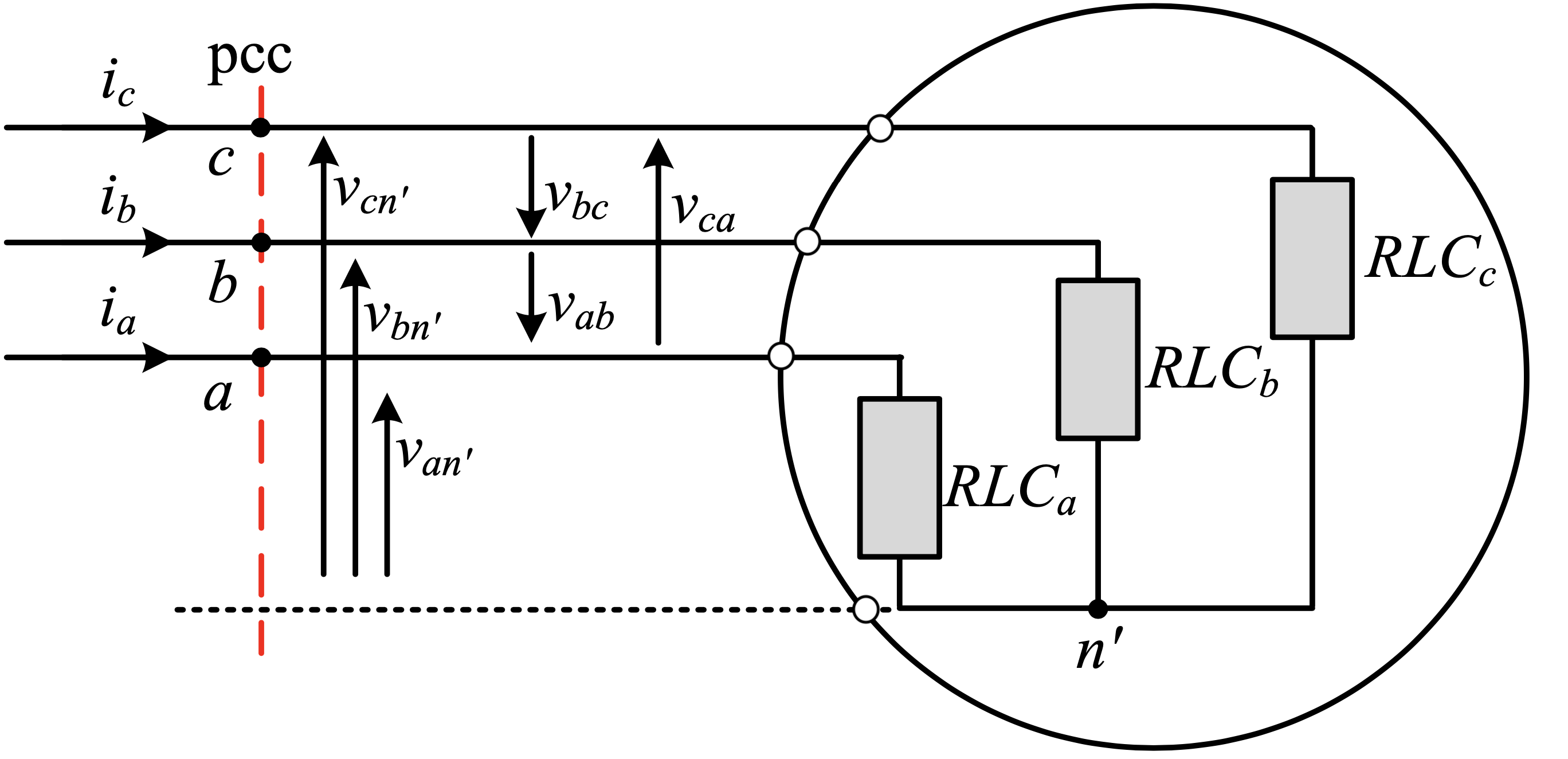}
	\caption{Virtual-neutral wye model for three-wire identification.}
	\label{fig:virtual_neutral_model}
\end{figure}
The virtual-neutral test uses a strongly unbalanced $RL$ branch with
$R_{ab}=0.306\,\Omega$ and $L_{ab}=0.7755\,\mathrm{mH}$, energized at
$\omega=200\pi$ rad/s by
\begin{align*}
	v_{ab}&=130\sqrt{2}\cos\omega t,\\
	v_{bc}&=120\sqrt{2}\cos(\omega t-2\pi/3),\\
	v_{ca}&=-(v_{ab}+v_{bc}).
\end{align*}
The identified virtual-wye sums recover the physical branch values:
$R_a+R_b=0.306\,\Omega$ and
$L_a+L_b=0.7755\,\mathrm{mH}$.
\begin{figure}[!t]
	\centering
	\includegraphics[width=\columnwidth]{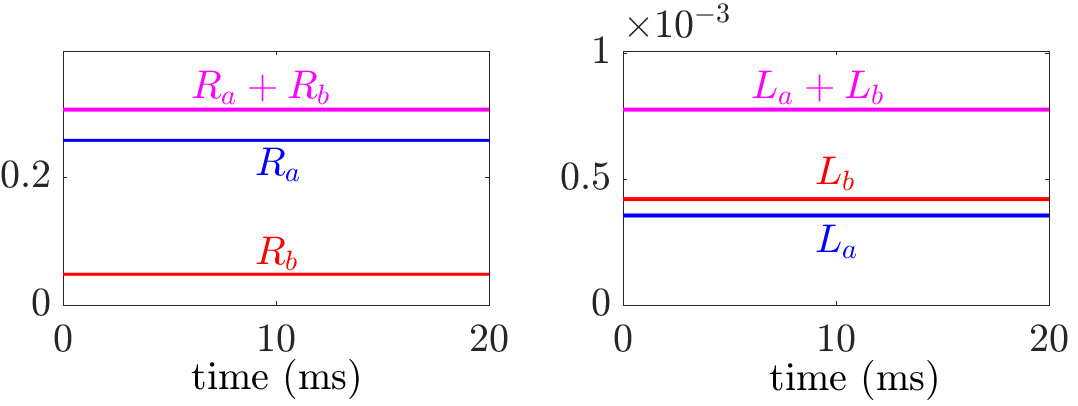}
	\caption{Virtual-neutral identification result. The key reported value is the recovered sum of virtual branch parameters.}
	\label{fig:virtual_neutral_identification}
\end{figure}
\subsection{Realistic MATLAB \newedit{Simulations}}
The preceding examples establish the algebraic structure of the proposed
three-wire equivalents. They do not, by themselves, answer the practical
question raised by sampled measurements: whether the physical equivalent can
still be recovered when the observer has only finite windows of noisy terminal
waveforms. The MATLAB \newedit{simulations} in this section address that question. Their
purpose is not to benchmark a generic system-identification routine or to force
an answer for every possible waveform; it is to test whether the method recovers
a correct \vtwo{load-equivalent model} when the measurement record contains
sufficient information, and whether the diagnostic quantities expose cases that
are poorly conditioned or biased by the measurement chain. \vtwo{This study is
simulation-only by design: it isolates the three-phase identifiability problem
under controlled, known-ground-truth conditions. Laboratory validation of the
single-phase formulation is reported in~\cite{arrabal2025experimental}, and
three-phase hardware validation is left to future work, but there is a high degree of confidence since single phase cases were properly validated with strenuous circumstances.}

Two sets of simulations are used. The first set is a topology-and-order
\newedit{study}, covering balanced and unbalanced delta parallel RLC and wye series
RLC loads, including an extreme-unbalance case and a deliberately
low-information single-frequency case. The second set is a measurement-chain
stress \newedit{study} applied to a fully unbalanced three-wire delta parallel RLC
load. This second case is intentionally demanding because all branch
resistances, inductances, and capacitances must be recovered from terminal
line-to-line voltages and line currents.

All simulated records are generated from a 50 Hz unbalanced multisine source
sampled at 20 kHz. The phase voltages include amplitude imbalance and low-order
harmonics,
\begin{equation}
	v_x(t)=\sum_{h\in\mathcal H_x} V_{x,h}
	\sin(h\omega_0 t+h\phi_x+\alpha_{x,h}),
	\qquad x\in\{a,b,c\}.
	\label{eq:realistic_multisine}
\end{equation}
so the estimator is not using a phasor model extracted from ideal sinusoids.
For every run, the input data are only the terminal waveforms available at the
three-wire point of connection. \newedit{Each sliding window spans 0.75 of a
fundamental cycle (15~ms, i.e.\ 300 samples at 20~kHz) and advances in steps of
0.10~cycle (2~ms); the windows are processed independently.}
For each window, the implementation computes the rank and condition number of
the regression matrix, KCL/KVL consistency, terminal power residual, energy
residual, passivity, and physical parameter estimates. The internal delta
parallel coordinates \(G,\varGamma,C\) are converted back to the reported physical
parameters \(R=1/G\), \(L=1/\varGamma\), and \(C\).

Table~\ref{tab:smoke_results} gives a compact baseline check. The clean cases
close the terminal power balance at machine precision, while the 60 dB cases
increase the residual to the level expected from the injected measurement
noise. This table is deliberately simple: it verifies that the sampled-data
implementation preserves the exact geometric construction before the more
demanding stress cases are considered.

\begin{figure*}[!t]
	\centering
	\includegraphics[width=0.85\textwidth]{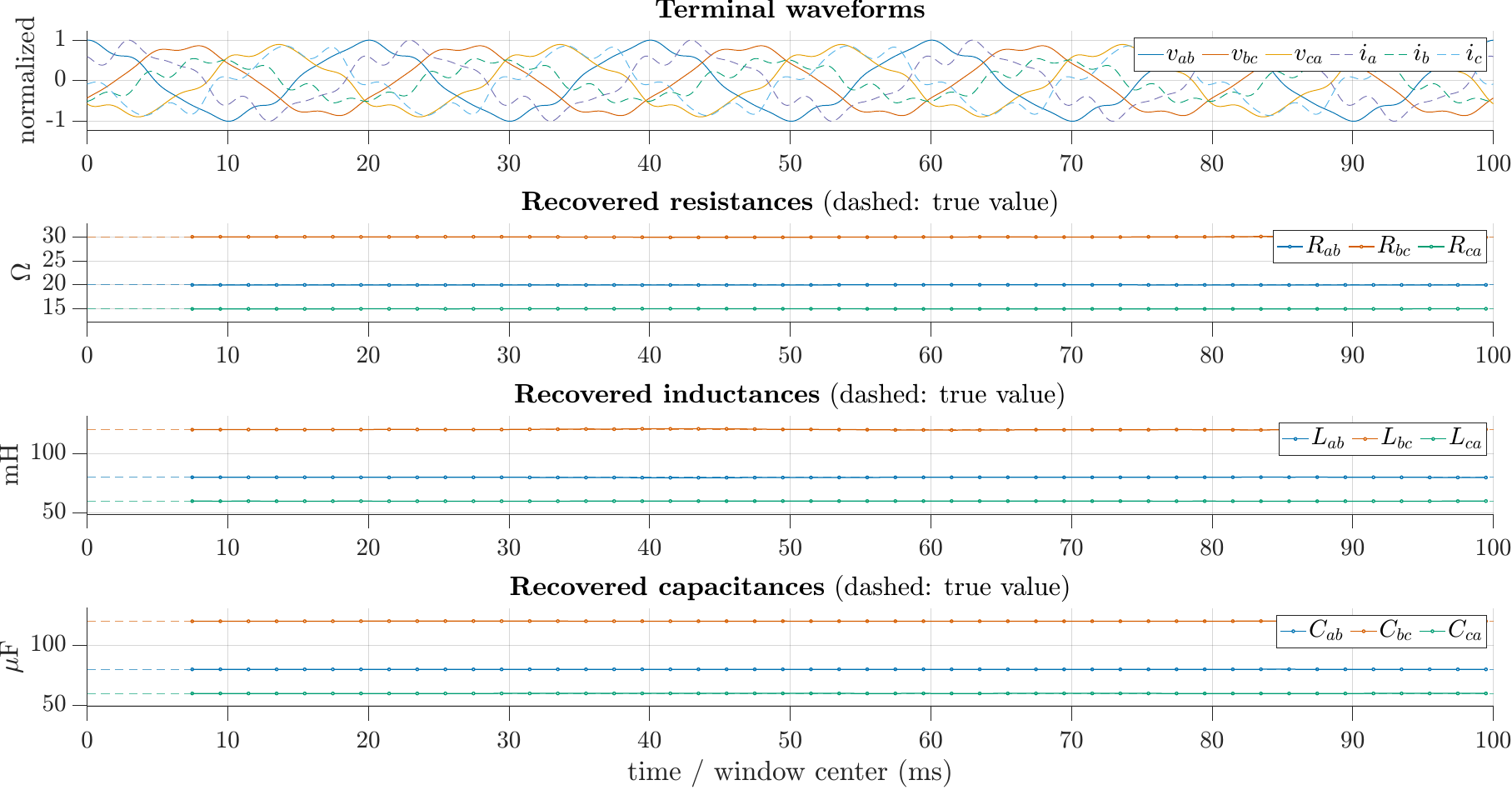}
	\caption{Representative three-wire delta parallel RLC identification under 70 dB
	measurement noise. The upper panel shows the measured terminal waveforms
	over five cycles. The lower panels show recovered resistances, inductances,
	and capacitances in physical units. Dashed lines indicate the
	known values used to synthesize the load.}
	\label{fig:delta_glc_noisy_recovery}
\end{figure*}

Fig.~\ref{fig:delta_glc_noisy_recovery} then shows the main reporting style for
one representative stress case. The upper panel displays five cycles of the
measured terminal voltages and currents. The lower panels show the recovered
branch \(R\), \(L\), and \(C\) values in engineering units for successive
windows; dashed lines are the known parameters used to synthesize the load.
This figure is important because it avoids hiding the method behind regression
coordinates: the reader sees the measured waveforms and the
\newblock{physical load-equivalent parameters} that are recovered from them.

\begin{figure*}[!t]
	\centering
	\includegraphics[width=0.85\textwidth]{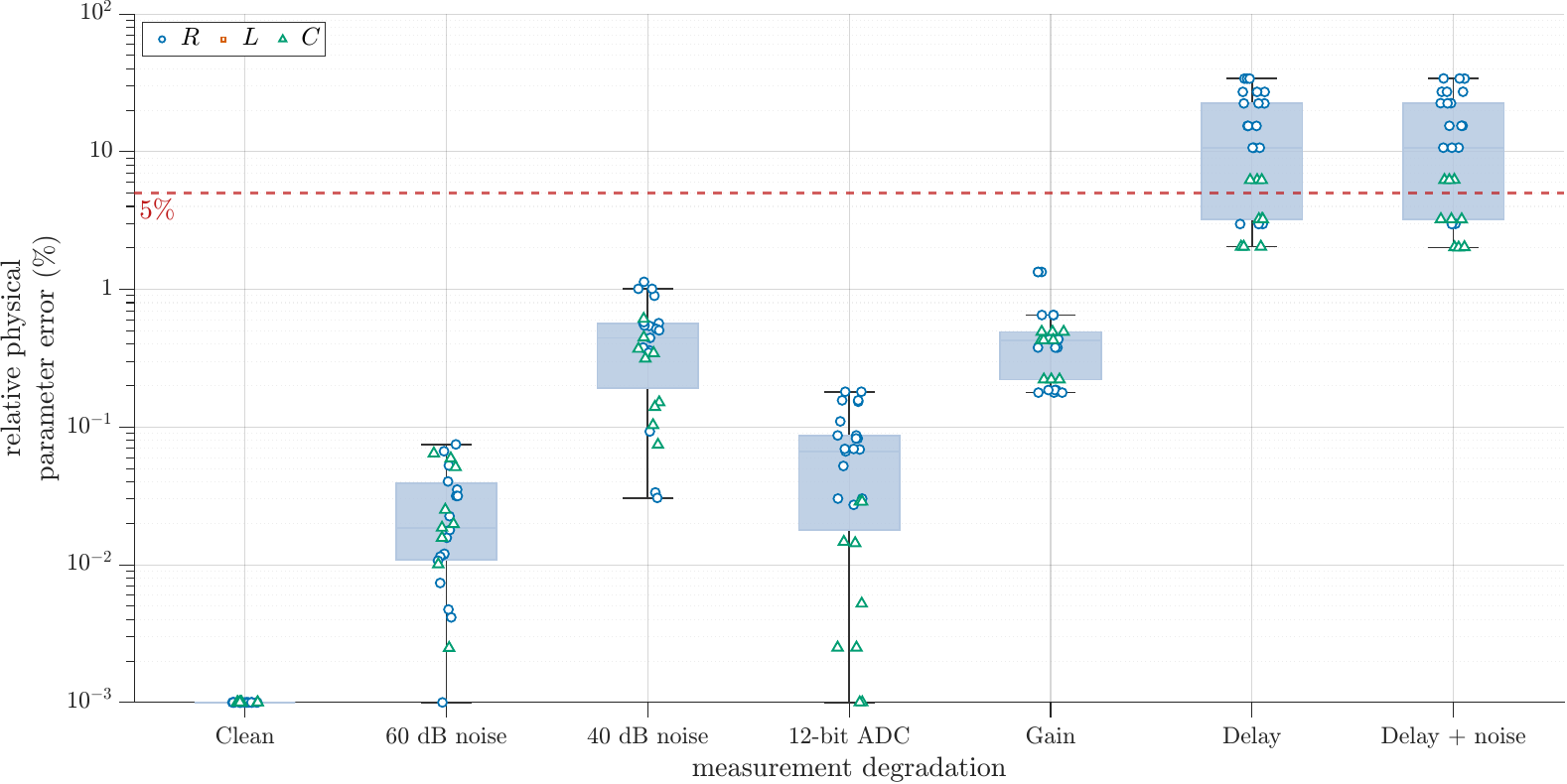}
	\caption{Measurement-chain stress test for a three-wire delta parallel RLC
	load. Boxes summarize the distribution of relative physical-parameter error
	across branch \(R\), \(L\), and \(C\) estimates and practical window lengths
	from 0.5 to 2 cycles; markers show the individual element classes. The
	dashed line marks 5\%. The deliberately fragile 0.25-cycle windows are
	retained in the reproducibility data set but excluded from this main-paper
	view.}
	\label{fig:degradation_error_boxes}
\end{figure*}

The measurement-chain stress \newedit{study} applies the perturbations that are most
likely to degrade this type of time-domain estimator: white measurement noise
at 60 dB and 40 dB SNR, 12-bit quantization, voltage/current gain mismatch, a
50 \(\mu\)s coherent current-sensor delay, and the combined delay-plus-noise
case. Practical window lengths from 0.5 to 2 cycles are shown in
Fig.~\ref{fig:degradation_error_boxes}; deliberately fragile 0.25-cycle windows
are retained in the reproducibility data set but excluded from the main figure.
The boxes summarize the distribution of relative errors over branch \(R\),
\(L\), and \(C\) estimates. Clean data, 60 dB noise, 40 dB noise, 12-bit
quantization, and gain mismatch remain within the 5\% engineering band in all
practical-window settings. Current-sensor delay is different: it produces a
coherent phase bias rather than random scatter, and the resulting parameter
errors are therefore correctly visible as a synchronization problem rather than
as an uncertainty that can be averaged away.

\begin{table}[!t]
\centering
\caption{Baseline sampled-data checks for the three-wire implementation.}
\label{tab:smoke_results}
\begin{tabular}{lccc}
\toprule
Case & Coverage & Median cond. & Power residual \\
\midrule
Delta clean & 1.00 & 21.83 & $5.30\times10^{-16}$\\
Delta 60 dB & 1.00 & 21.83 & $1.51\times10^{-3}$\\
Wye clean & 1.00 & 3.52 & $1.61\times10^{-14}$\\
Wye 60 dB & 1.00 & 3.52 & $2.95\times10^{-3}$\\
\bottomrule
\end{tabular}
\end{table}

\begin{table*}[!t]
\centering
\caption{Representative behavior of the realistic three-wire MATLAB \newedit{study}.}
\label{tab:realistic_campaign}
\begin{tabularx}{\textwidth}{@{}l X c@{}}
\toprule
Scenario & Main observation & Information content\\
\midrule
Pure fundamental & The trajectory is rank deficient for higher-order
three-wire RLC models. Noise may create numerical rank but not physical
information. & Insufficient\\
Weak harmonics, clean & Parameters are recovered when the window is at least
0.5 cycles; shorter windows are fragile. & Sufficient\\
Weak harmonics, 60 dB & Residuals remain compatible with injected noise for
0.5--2 cycle windows. & Sufficient\\
Weak harmonics, 40 dB & Longer windows, typically 1--2 cycles, are required for
stable estimates. & Conditional\\
Rich harmonics & Conditioning improves markedly; median condition is about 22
for delta parallel and 3.5 for wye series in the tested cases. & Sufficient\\
50 $\mu$s current delay & The error is coherent rather than random; delta
parallel RLC reaches about 37\% maximum parameter error in the stress test,
despite otherwise acceptable rank and coverage. & Biased\\
\bottomrule
\end{tabularx}
\end{table*}

The same observations are summarized in Table~\ref{tab:realistic_campaign}. It
separates three situations that must not be conflated. First, low-rank
waveforms, such as a pure fundamental excitation for a high-order RLC
equivalent, are not informative enough to identify all stored-energy
coordinates. Second, additive noise and quantization mainly broaden the error
distribution; their effect is reduced by informative windows of practical
length. Third, coherent timing error is a bias in the measured relationship
between voltage and current; it must be calibrated or diagnosed, not treated as
zero-mean noise.

For the degradation study in Fig.~\ref{fig:degradation_error_boxes}, a
configuration is considered high-confidence when
\begin{equation}
\begin{aligned}
\mathrm{coverage}&\ge 80\%,&
\mathrm{max\;relative\;error}&\le 5\%,\\
\rho_E&\le 5\%,&
\kappa&\le 10^6 .
\end{aligned}
\label{eq:high_confidence_gate}
\end{equation}
Fifteen of the 21 practical-window configurations in the degradation \newedit{study}
satisfy this gate. The six configurations that do not satisfy it are precisely
the current-delay and delay-plus-noise cases. This is the desired behavior for
a field method: the estimator should recover physical parameters under
ordinary measurement noise and quantization, but it should not disguise a
coherent synchronization error as a reliable \newblock{load equivalent}.

\newblock{The additional derivations, expanded \(s_I\) coefficients, nine-case
topology/operating-condition summaries, degradation heat maps, window
diagnostics, extra waveform-parameter examples, and MATLAB reproduction scripts
are maintained as online material at \texttt{https://electrica.ual.es/spacor}.
The printed paper keeps only the figures needed to support the principal claims.}

\section{Discussion}

The method identifies an equivalent only when the measured trajectory contains
the information required by the candidate model. This is essential for field
use. An algorithm that always returns parameters can be numerically convenient
but physically misleading. The proposed diagnostics make the result auditable:
one can see whether the model is passive, whether Kirchhoff constraints are
satisfied, whether the energy-admissibility residual is small, and whether the
local geometry is sufficiently rich. This is more than an error metric: it is
the physical test that connects parameter identification with the energy
equivalent \vtwo{for the assumed topology} \newblock{reported in the same
observation window}.

\vtwo{When several admissible topologies pass the rank, conditioning, and
energy-admissibility tests, they are terminally indistinguishable from the given
window: each closes the same terminal energy balance while assigning the stored
and dissipated energy to different branches. One then reports the admissible
alternatives or resorts to a prior topology or additional excitation, since no
selection rule exists at the level of terminal data alone; characterizing that
non-uniqueness is a separate, energy-level question.}

The present manuscript also clarifies the relation between exact geometric
formulae and practical sampled-data implementation. The symbolic quotients
establish the analytical structure of the identifiable models, while the
windowed matrix form is the implementation used to address noise, derivatives,
conditioning, and computational feasibility.

Compared with event-driven aggregate load identification and other
time-varying parameter approaches \cite{renmu2006composite,wang2017robust},
the proposed estimator is narrower but more physically structured: it identifies
the parameters of a selected \newblock{load equivalent} directly from local terminal
waveforms. It should therefore be judged by whether the selected topology is
identifiable, passive, and energetically consistent in the observation window,
not by whether it can emulate arbitrary black-box dynamics.

\section{Conclusions}

This paper identifies three-phase load equivalents from terminal measurements
in the time domain, with particular attention to the three-wire case. The
implementation adds the practical layer required for sampled data: finite
windows, normalization, explicit identifiability tests, measurement-degradation
studies, and energy-admissibility validation. \newedit{The practical value is
that the result is an equivalent circuit with the same energetic behavior as the
measured load: energy is consumed and stored or restored in the same fashion as
in the physical circuit.} The identified equivalents therefore provide
the physical state required for \newblock{element-level energy accounting within
the recovered load model and for monitoring, protection, or compensation
applications.}

\newedit{Several applications follow from this capability. Because the
equivalent is recovered from local windows of terminal data and needs no neutral
connection, it suits three-wire distribution feeders, motor and delta-connected
loads, and converter interfaces where only line quantities are accessible.
Tracking the recovered branch \(R\), \(L\), and \(C\) across windows turns the
method into a condition-monitoring tool, with the energy-admissibility and
conditioning diagnostics separating genuine parameter drift from measurement
artefacts such as the coherent sensor delay reported above. The same physically
meaningful equivalent supports impedance-based and adaptive converter control,
relay setting and anomaly detection in protection, and the sizing of
compensation hardware. In all of these the confidence gate is decisive for field
use: the estimator either returns an auditable equivalent or declares the window
uninformative, rather than a value that cannot be trusted.}

\appendices
\section{Windowed Matrix and Harmonic-Projection Implementation}
\label{app:windowed_implementation}

Let \(z[k]\), \(k=0,\ldots,N-1\), denote any measured terminal waveform in one
local window. The implementation first builds the signal coordinates required by
the selected physical equivalent. For example, a delta parallel RLC branch uses
\(v_{xy}\), \(\breve v_{xy}\), and \(v_{xy}'\), whereas a wye series RLC
branch uses current coordinates and their corresponding derivatives or
primitives. Stacking all samples in the window gives
\begin{equation}
	\mathbf y_w=\Phi_w\theta+\epsilon_w,
	\label{eq:appendix_window_problem}
\end{equation}
where \(w\) denotes the window index. Before solving, the columns and output are
scaled as
\begin{equation}
	\tilde{\Phi}_w=\mathbf W_y\Phi_w\mathbf W_\theta^{-1},
	\qquad
	\tilde{\mathbf y}_w=\mathbf W_y\mathbf y_w ,
	\label{eq:appendix_scaling}
\end{equation}
with diagonal scale matrices computed from column and output norms. The
least-squares estimate is obtained from the scaled system and then mapped back
to physical units:
\begin{equation}
	\hat\theta_w=\mathbf W_\theta^{-1}
	\tilde{\Phi}_w^{\dagger}\tilde{\mathbf y}_w .
	\label{eq:appendix_scaled_ls}
\end{equation}
The reported condition number is the condition of the scaled problem, so it
measures information content rather than the arbitrary choice of volts, amperes,
henries, or farads.

The harmonic-projection derivative used in the multisine \newedit{simulations} is a
window-local model, not a frequency-domain power decomposition. A measured
signal is approximated by
\begin{equation}
	z(t)\approx c_0+\sum_{h\in\mathcal H_w}
	\left[a_h\cos(h\omega_0 t)+b_h\sin(h\omega_0 t)\right],
	\label{eq:appendix_harmonic_projection}
\end{equation}
where \(\mathcal H_w\) is either prescribed or selected from candidate
harmonics using relative amplitude, energy fraction, sample budget, and basis
conditioning. Differentiation is then analytical:
\begin{equation}
\begin{aligned}
	\frac{d^q z}{dt^q}\approx
	\sum_{h\in\mathcal H_w}(h\omega_0)^q
	\Big[
	&a_h\cos\!\left(h\omega_0 t+\frac{q\pi}{2}\right)\\
	&+b_h\sin\!\left(h\omega_0 t+\frac{q\pi}{2}\right)
	\Big].
\end{aligned}
	\label{eq:appendix_harmonic_derivative}
\end{equation}
For primitive coordinates, the same fitted harmonics are integrated
analytically after removing the local mean; when a direct integral form is
available, it is preferred over a differentiated form because it is less
sensitive to quantization.

Each window produces not only \(\hat\theta_w\), but also a health record:
\begin{equation}
	\mathcal D_w=\{\operatorname{rank}(\tilde{\Phi}_w),\,
	\kappa(\tilde{\Phi}_w),\,\rho_{\mathrm{KCL/KVL}},\,
	\rho_p,\,\rho_E,\,\Pi\},
	\label{eq:appendix_diagnostics}
\end{equation}
where \(\rho_{\mathrm{KCL/KVL}}\) is the equation residual, \(\rho_p\) is the
terminal-power residual, \(\rho_E\) is the energy-admissibility residual defined
in \eqref{eq:energy_admissibility_residual}, and \(\Pi\) records passivity of
the recovered elements. This diagnostic record is what allows the method to
distinguish an informative noisy window from a low-rank or synchronization-biased
one.

\bibliographystyle{IEEEtran}
\bibliography{rebuild_references}

\end{document}